# K-SAM: A Prompting Method Using Pretrained U-Net to Improve Zero Shot Performance of SAM on Lung Segmentation in CXR Images

Mohamed Deriche, Mohammad Marufur

**Abstract**:

**Background and Objective**: In clinical procedures, precise localization of the target area is an essential step for clinical diagnosis and screening. For many diagnostic applications, lung segmentation of chest X-ray (CXR) images is an essential first step that significantly reduces the image size to speed up the subsequent analysis. One of the primary difficulties with this task is segmenting the lung regions covered by dense abnormalities also known as opacities due to diseases like pneumonia and tuberculosis. SAM has astonishing generalization capabilities for category agnostic segmentation. In this study we propose an algorithm to improve zero shot performance of SAM on lung region segmentation task by automatic prompt selection.

**Methods**: Two separate U-Net models were trained, one for predicting lung segments and another for heart segment. Though these predictions lack fine details around the edges, they provide positive and negative points as prompt for SAM.

**Results**: Using proposed prompting method zero shot performance of SAM is evaluated on two benchmark datasets (Montgomery and Shenzhen). ViT-l version of the model achieved slightly better performance compared to other two versions (ViT-h and ViT-b). It yields an average Dice score of 95.5% and 94.9% on hold out data for two datasets respectively. Though, for most of the images, SAM did outstanding segmentation, its prediction was way off for some of the images. After careful inspection it is found that all of these images either had extreme abnormality or distorted shape.

**Conclusion**: Unlike most of the research performed so far on lung segmentation from CXR images using SAM, this study proposes a fully automated prompt selection process only from the input image. Our finding indicates that using pretrained models for prompt selection can utilize SAM's impressive generalization capability to its full extent.

Keywords: Lung segmentation, SAM, UNet, zero-shot learning, medical imaging.

## I. Introduction:

Creation of functional and visual representations of the interior of the human body is crucial for medical analysis. Medical imaging makes it possible to perform diagnostic procedures in non-invasive methods.

Due to the advent of technology, high-resolution image acquisition has become easy and cost efficient for developing automated biomedical image analysis algorithms. Among various imaging modalities, Chest X-ray (CXR) is the most common and easily available technique used for diagnosis of thorax diseases because it is cheap and uses less radiation. For most of the computer-aided diagnosis systems to detect, identify, and analyze anatomical structures, lung segmentation of CXR images is an essential first step.

Segmentation is a basic step for automated analysis that subdivides the visually distinct regions with semantic information from the image. Extraction of precise borders or outlines of anatomical structures, tumors, lesions, or other areas of clinical importance is the objective of biomedical image segmentation. In many medical applications, such as disease diagnosis, therapy planning, surgical guidance, and disease progression tracking, accurate segmentation is essential. It enables medical professionals to measure volumes, track changes over time, analyze and quantify certain locations, and support them in making well-informed decisions regarding patient care. Moreover, segmentation facilitates further classification tasks by forcing the classification model to use only the relevant information from the segmented area, which in turn increases model's quality and reliability [1]. However, medical image segmentation is a challenging task due to low contrast, varied intensity and noisy images. Different shapes and sizes of anatomical organs across patients further complicate the segmentation process. Additionally, at the time of image acquisition unwanted artifacts and distortion may occur that significantly downgrade segmentation performance. Many segmentation strategies, from conventional techniques to more sophisticated deep learning algorithms, have been developed to address these issues. Image segmentation techniques can be categorized into three groups: 1) Manual segmentation, 2) Semi-automated segmentation, 3) Fully automated segmentation [2]. Manual segmentation involves subject experts to draw precise boundaries surrounding the region of interest which is used as the ground truth label for the development of semi-automated and fully automated segmentation techniques. Semi-automated segmentation techniques require user involvement with the algorithm in order to produce accurate segmentation results. On the other hand, fully automated techniques do not require any kind of user involvement. Such techniques are mostly based on supervised learning approaches that require labeled training data. Manual segmentation is a labor-intensive and time-consuming process, which can produce intra and inter observer variability. Fully automated computer aided lung segmentation is a crucial step for diagnosing lung diseases and giving proper treatment.

Deep learning (DL) based solutions achieved state-of-the-art performance in various biomedical applications including fully automated medical image segmentation [3] however, there remains opportunity for further improvements. DL algorithms require very large imaging datasets, which provoke the necessity of huge memory and computational power requirements. In order to annotate this large dataset a large number of domain experts is also needed, which is scarce in the medical domain.

In addition to these issues, a lot of imaging data are not publicly shared due to privacy and proprietary reasons. For biomedical image analysis, Convolutional Neural Network (CNN) has been the de-facto standard. The network could handle a diverse range of datasets and applications by stacking up multiple convolutional layers that includes object detection, classification, image registration, and segmentation. U-Net, which is a convolutional neural network utilizing encoder-decoder architecture, demonstrated unprecedented performance in biomedical image segmentation [4]. Various modified versions of U-Net architecture such as UNet++ [5], UNet 3+ [6], ResUNet++ [7] are being proposed to achieve state-of-the art performance.

After the advent of transformers that can capture long-range dependencies, Vision Transformer (ViT) based methods achieved remarkable performance in semantic segmentation [8]. However, the adaptability and generalization of these models are limited by the diversity and complexity of the images. To address this issue researchers and big tech organizations are introducing foundation models with remarkable generalization capabilities. A foundation model is usually trained with massive amount of data from several distinct distributions and can be used as a base for specialized tasks. Meta has recently introduced a foundation model called SAM with remarkable generalization capabilities. It was trained on the largest segmentation dataset containing over 1 billion masks for 11 million images. It was designed to be promptable so that it can transfer zero-shot to completely new image distribution and tasks. Due to the requirement of a prompt, using it for automated semantic segmentation task becomes challenging since its segmentation output highly depends on the prompt quality. Hence designing prompt selection method for semantic segmentation is a crucial task since [9].

Due to the challenges posed by the traditional methods and resource intensive characteristics of these methods for medical image segmentation, this study aims to examine efficiency of SAM on segmentation task. The contributions of this study are listed below:
1. Introduced a prompting method for SAM for lung region segmentation from CXR images.
2. Summarized the recent research on lung lesion segmentation from CXR images using SAM.
3. Compared zero shot performance of SAM on benchmark datasets with state-of-the-art deep learning models trained on specific data distribution.

The remainder of this paper is organized as follows: Section II provided summary of the recent state-of-the-art research, their pros and cons. Section III describes the details of the proposed method and required background information. Strengths and weakness of the proposed method is discussed based on the experimental results in section IV. Finally, section V draws conclusion of the paper.

## II. Related Works:

Medical image segmentation is a crucial research area, demanding adaptable models for precise identification of pathological regions and anatomical structures. Lung lesion segmentation from CXR images is an active research domain for researchers around the globe. While deep learning methods have demonstrated significant potential for enhancing lung segmentation accuracy and efficiency, foundation models like SAM are transforming this domain.

Most lung segmentation methods resize high resolution CXR images to lower resolution through linear interpolation for reducing computational complexity of deep neural network. Which in turn introduces boundary information loss and causes blurry boundary. Furthermore, for practical medical applications high-resolution images are necessary. When segmentation results obtained from downsampled images is upsampled the quality of segmentation is poor compared to original high-resolution images. Lee et al. [10] proposed an encoder-decoder CNN architecture with superpixel resizing framework in order to alleviate blurred boundaries introduced as a result of downsampling and upsampling. Sulaiman et al. [11] proposed a convolutional neural network architecture for lung segmentation from CXR images. Authors employed transpose layer in the concatenate block to improve the spatial resolution of the feature matrices generated by the previous layers. They claimed to achieve state-of-the-art performance on a dataset collected from Kaggle that outperformed some improved UNet models such as ResUNet and VGGUNet. Another study [12] used UNet, UNet++, Attention UNet, Attention UNet++, and PSP Attention UNet to segment two types of tuberculosis (TB) lesions in CXR images. Final output was generated by applying ensemble techniques on the output form top five models.

Liu et al. [13] improved the UNet architecture using pre-trained Efficientnet-b4 as the encoder and incorporated LeakyReLU activation function and Residual block in the decoder. The improved model outperformed conventional UNet model in lung segmentation task and showed 97.92% and 97.82% dice score on two benchmark datasets (JSTR and Montgomery Country) respectively. A two-stage workflow for segmenting lung region in the CXR image and scoring severity of COVID-19 were proposed in this study [14]. Nine state-of-the-art segmentation models were trained and performance were compared in this study for both stages. DeepLabV3+ and MA-Net performed significantly better than other models for lung segmentation and diseases segmentation respectively. Finally, in the postprocessing step using lung and diseases segments a severity score were measured. Souza et al. [15] proposed an unique method which incorporated two CNN models: one for initial segmentation and another for segmentation reconstruction. The initial segmentation model is based on AlexNet, that takes a ($32 \times 32$) patch from the CXR image and predicts whether the patch is from lung or not. Then lung patches were plotted and morphological operations were applied to remove false positive patches. In case of healthy patients this initial segmentation produced great results. The reconstruction model utilized ResNet18 architecture and toke initial segmentation as input to produce a reconstructed mask.

The final lung segment was produced by combining these two into a single mask. The method was evaluated on the MC dataset and obtained 94% dice score.

Another study [16] proposed a context-aware deep fused CNN-Transformer architecture with semi-supervised global and spatial attention mechanisms. It aims to combine strengths of both CNN and Vision Transformer (ViT) architectures and complement their limitations. The feature tensors obtained from two branches (CNN, ViT) were fused using the proposed attention mechanism and a hybrid loss function for better adaptation. Researchers observed outstanding performance of polyp segmentation tasks in multiple data sources. Lyu and Tian [17] proposed U-shaped multiple tasking Wasserstein generative adversarial network (MWG-UNet), that takes advantage of attention mechanism to improve segmentation accuracy of the generator. Improved U-Net architecture called ARU-Net with squeeze and excitation (SE) blocks was used as the generator and several convolution layers along with fully connected layers as the discriminator. The generator produced segmentation masks which is passed to the discriminator along with the ground truth segmentation masks. Though authors achieved high dice score, their method showed sub-optimal IoU score.

Mazurowski et al. [18] extensively evaluated SAM's performance on 19 medical anatomical imaging datasets collected from different modalities. They observed that SAM's performance fluctuates based on the image modality. SAM did well in segmenting ilium from X-ray images and on the other hand did poorly on segmenting gray matter from MRI images. They concluded that, though SAM has potential to make impact in medical image segmentation, proper precautions need to be taken when using it. There are very limited works that utilized SAM particularly on lung lesion segmentation from CXR images. One of such study [19] showcases SAM's robustness in segmenting lung lesions from Computed Tomography (CT) and CXR images. As prompt, a bounding box containing the lung masks is taken from the user and some positive and negative points were selected from the mask. Centroid of each lung mask is selected as positive points, and the center point of the lung bounding box as negative point. Though their approach achieved 93.19% Dice score, it has limitations in the process. The process require bounding box from the user and positive and negative points were calculated from the mask which should not be available in inference time. Another study [20] presents prompt integrated framework for automated lung segmentation from CXR images. Multiple points on the lung area were taken from the user and passed as the prompt. Khalili et al. [21] explored performance of a fully automated framework with the You Only Look Once (YOLO) model for providing prompts to the SAM. Author observed that geometric prompt has positive effect on SAM's generalization capabilities. The YOLO model was trained to predict bunding boxes around the lung segments. These bounding boxes along with the mask was passed to the mask encoder. Authors reported to have achieved 97.16% and 95.25% Dice factor on Montgomery and Shenzhen dataset respectively. Since mask is passed in the prompt encoder this score may not properly demonstrate SAM's performance.

## III. Methods

*A. Datasets:*

In this study two publicly available pre-processed benchmark datasets for lung segmentation were collected (Montgomery and Shenzhen) [22]. These datasets contain CXR images collected from patients diagnosed with either tuberculosis, pneumonia, or COVID-19. These datasets were manually annotated with masks excluding the heart region by expert radiologists. The Montgomery dataset contains 139 CXR images, out of which 80 images are of normal category, and the remaining images diseased category. The Shenzhen dataset have 566 CXR images. For the requirement of this study, we also annotated CXR images of both datasets with their corresponding heart segmentation mask. For annotation purpose Computer Vision Annotation Tool (CVAT) [23] were used. Figure 1 shows chest x-ray images and their ground truth masks from both datasets.

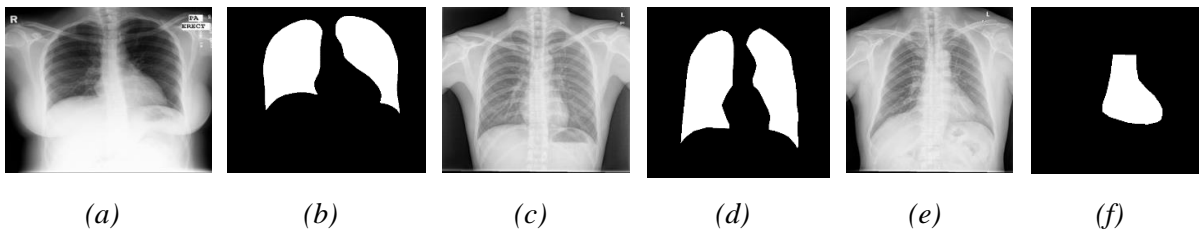

*(a)*  *(b)*  *(c)*  *(d)*  *(e)*  *(f)*

Fig. 1. Chest X-ray images and corresponding masks.

*B. Data Preprocessing:*

All the CXR images in the dataset were in RGB format with shape $(512 \times 512 \times 3)$ and corresponding masks were of shape $(521 \times 512)$. For input of the SAM, unprocessed original image preserving the size and quality were used. To train UNet model for preliminary segmentation model for lung segmentation and heart segmentation tasks both datasets were preprocessed. Images were converted to gray scale and resized to $(128 \times 128 \times 1)$ shape and masks to $(128 \times 128)$ in order to reduce computation cost. Furthermore, pixel intensity values were normalized using min-max normalization technique. Both datasets were splitted into three parts: 70% used for training purpose, 10% for validation, and remaining 20% hold out for testing purpose.

*C. U-NeT with Pre-Trained Encoder:*

Conventional U-Net architecture with encoder and decoder structure is widely utilized for biomedical image segmentation. U-Net architecture combines low-level features maps with the higher-level feature maps using skip connections. While training a U-Net model with randomly initialized weights for specific tasks having small datasets is prone to overfit. Instead of training from scratch using a pre-trained encoder base can improve the performance of U-Net architecture [24].

D. *K-Medoids Clustering Algorithm:*

K-medoids clustering is a variant of K-means clustering algorithm. The most centrally located object of a cluster is called the medoid. The medoids that are actual points from the data represents the clusters. In K-means clustering algorithm the cluster representor is the centroid which is calculated by averaging all the cluster data points. Due to the difference in methodology K-means clustering algorithm is more sensitive to outliers and easily influenced by extreme values than K-medoids clustering algorithm. Figure 2 depicts difference between the K-medoids and K-means algorithm in a 2-D example.

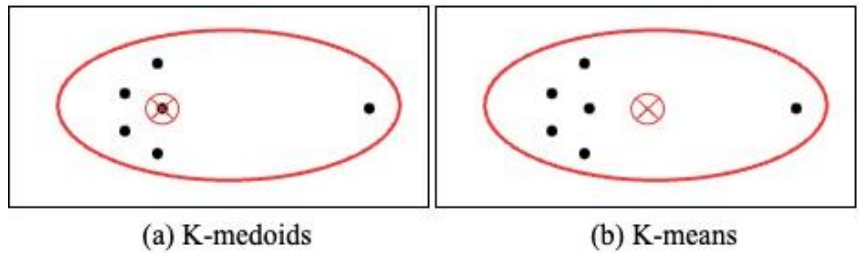

Fig. 2. K-means vs K-medoids in 2-D plane. In both figures (a) and (b), the points in the left form a cluster and the rightmost point is an outlier. The red point represents the cluster center.

E. *SAM:*

The Segment Anything Model (SAM) is a promptable segmentation model proposed by Meta AI Research in their Segment Anything project as foundation model for image segmentation [25]. The SAM is an encoder-decoder based architecture that contains three key components: an image encoder, a prompt encoder, and a mask decoder. Self-supervised pre-trained Vision Transformer (ViT) was utilized as the image encoder with minimal adaptation to high resolution images. SAM is equipped with three different ViT backbone of varied sizes: ViT-h (Huge), ViT-l (Large), ViT-b (Base). For prompt encoders, two types of prompts are considered: dense (masks) and sparse (text, points, boxes). The mask decoder maps the image embedding, prompt embedding, and an output token to a mask. It uses prompt self-attention and cross-attention from prompt to image embedding and vice-versa to update both embeddings. SAM was trained with the largest segmentation dataset containing over 1 billion masks for 11 million images. Since, the dataset did not include any medical images, it makes it perfect for using the aforementioned lung segmentation datasets to assess its zero-shot transfer capabilities.

F. *K-SAM Proposed Method*

The proposed architecture (K-SAM) which is shown in figure 3. SAM takes two inputs: the input CXR image and the prompt. As input image to the SAM, original image of shape $(512 \times 512 \times 3)$ was passed directly without any processing. For selecting appropriate prompt for the input CXR image it was pre-processed and passed through the prompt selection process.

Two separately trained U-Net models take processed CXR image and generates lung and heart masks. This predicted masks though represents generic shape of the lung and heart regions, lacks fine details. These masks are used to select total ten points: two positive points at the center of lung segments, five negative points outside the lung region, and three negative points inside the heart region. In this study, we aim to predict the lung segment excluding the heart area. So, points in the heart region are selected as negative points. These ten points were chosen so that SAM only predicts lung regions excluding the heart region. In order to select these points K-medoids clustering algorithm was used. In order to select the positive points, two cluster centers (medoids) for the pixels inside two lung segments (left and right) were measured. Similarly, five cluster center were measured for the pixels outside the lung segments regions and three from the pixels in heart region. Figure 4 shows selection of positive and negative points as prompts for the SAM model using K-medoids algorithm.

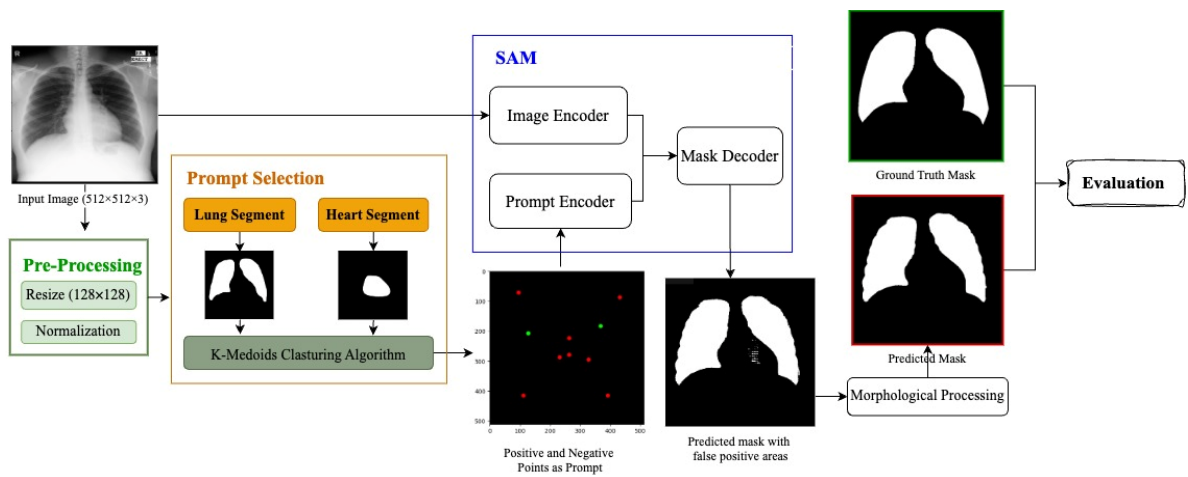

Fig.3. Block diagram of the proposed segmentation framework.

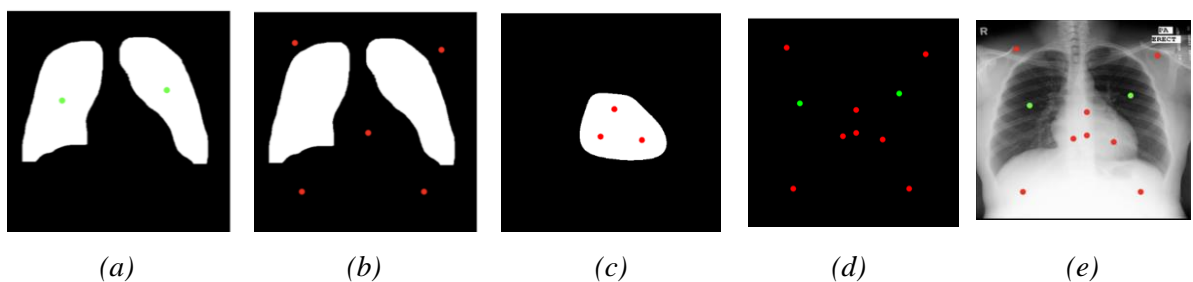

*(a)* *(b)* *(c)* *(d)* *(e)*

Fig. 4. Prompt selection using K-medoids clustering algorithm. (a) Selection of two cluster centers in the lung region as positive points, (b) Five cluster centers outside the lung region as negative points, (c) Three negative points in the heart area, (d) Reference points as prompt for SAM model, (e) Reference points are plotted on the input CXR image.

The immediate output of the SAM contained some small misclassified regions that was completely removed by post-processing steps. In post-processing two morphological operations erosion and dilation were performed. At first erosion was executed for three iterations which was followed by dilation for three iterations. The number of iterations was considered three for both operations which was selected by trial-and-error process. Figure 5 shows input image and output of SAM before and after post-processing.

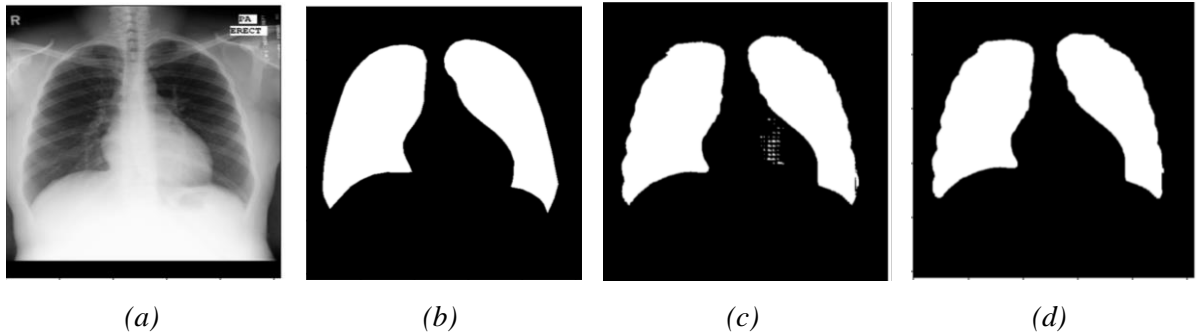

*(a)* *(b)* *(c)* *(d)*

Fig.5. Post-processing of the output form SAM. (a) Input chest x-ray image for SAM, (b) Ground truth mask, (c) Predicted mask from the SAM, (d) Final mask after morphological processing.

## G. Performance Metrices

To assess how accurate the predicted mask is Dice Coefficient, Intersection-Over-Union (IoU), and Cohen's Kappa score were calculated. These matrices are used in state-of-the-art research in medical images segmentation. The pixels of the predicted mask can be grouped into four groups: True Positive (TP), False Positive (FP), True Negative (TN), and False Negative (FN). Equation (1) for calculating dice score, (2) for IoU score, and (3)-(8) were used for calculating Cohen's Kappa score.

$$Dice = \frac{2TP}{2TP + FP + FN} \quad (1)$$

$$IoU = \frac{TP}{TP + FP + FN} \quad (2)$$

$$Cohen's\ Kappa, K = \frac{P_o - P_e}{1 - P_e} \qquad (3)$$

$$P_O = \frac{TP + TN}{Total} \qquad (4)$$

$$P_e = P_{Correct} + P_{Incorrect} \qquad (5)$$

$$P_{Correct} = \frac{TP + FN}{Total} \times \frac{TP + FP}{Total} \qquad (6)$$

$$P_{Incorrect} = \frac{FP + TN}{Total} \times \frac{FN + TN}{Total} \qquad (7)$$

$$Total = TP + FP + TN + FN \qquad (8)$$

## IV. Result Analysis and Discussion:

The prompt selection process consists of two steps: first step is to train preliminary heart and lung segmentation models and second step is using k-medoids algorithm calculating reference points. In this study, multiple U-Net architecture with different encoder architecture pre-trained on ImageNet dataset was trained using 70% data from Montgomery and Shenzhen datasets. Among five encoders (VGG16, VGG19, Xception, ResNet34, and DenseNet169), U-Net with VGG19 encoder performed relatively well for both lung and heart segmentation tasks. For lung segmentation Dice score, IoU score, and Cohen's Kappa score respectively 0.88, 0.86, 0.84 were observed and 0.882, 0.889, and 0.874 for heart segmentation. The model was trained for 20 epochs and stopped early due to no significant change in the validation performance in both cases. In figure 6 and figure 7 performance of the trained segmentation model on images from hold out dataset is shown.

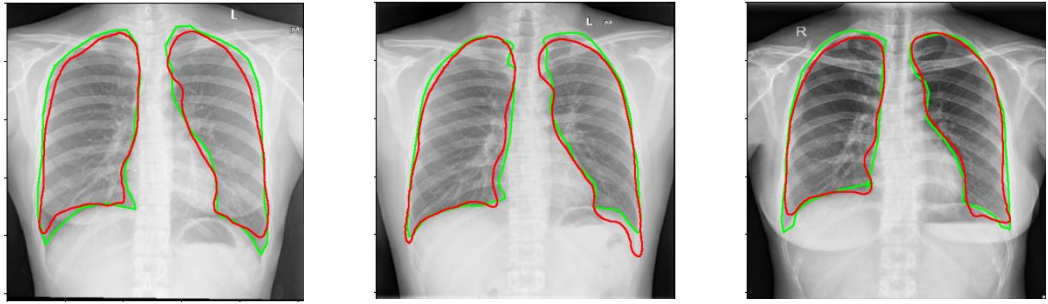

*Fig. 6. Performance of the U-Net with VGG19 Pre-trained encoder architecture in lung segment prediction. Green and red regions are ground truth and predicted lung segments respectively.*

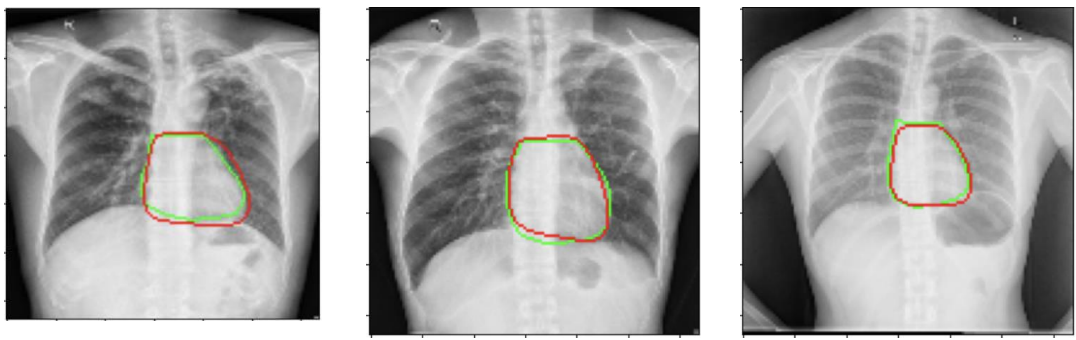

*Fig. 7. Performance of the U-Net with VGG19 Pre-trained encoder architecture in heart segment prediction. Green and red regions are ground truth and predicted heart segments respectively.*

Both k-means and k-medoids clustering algorithms were used for selecting reference points as prompt. Prompts selected by K-medoids algorithm showed significantly high-performance score compared to K-means algorithm. Performance of these algorithms is depicted in figure 8. When K-means clustering algorithm was used in some cases unwanted regions were selected as reference point. One of such case is shown in figure 5(b), it is found that a point inside the lung region is selected as negative point by the K-means algorithm which is not the case for K-medoids algorithm.

SAM showed impressive zero-shot performance that outperformed our U-Net architecture with VGG19

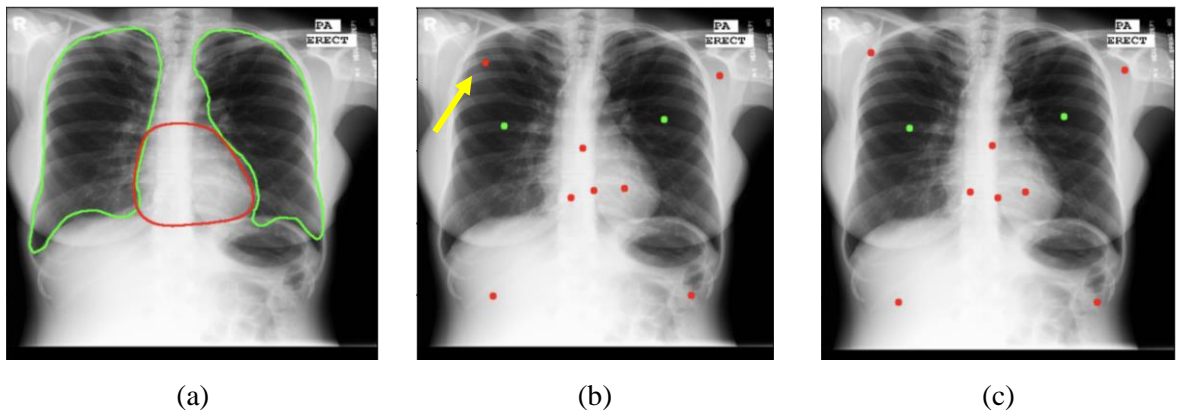

(a) (b) (c)

Fig. 8. Difference between performance of K-means and K-medoids clustering algorithm. (a) X-ray image with annotated lung (green) and heart (red) region, (b) Positive points (green) and negative points (red) selected as prompt when K-means algorithm is used, (c) Positive points (green) and negative points (red) selected as prompt when K-medoids algorithm is used.

encoder. The U-Net model for lung segmentation was trained using CXR data for specific task and evaluated using data from similar distribution. Since SAM did not see any CXR image at the time of training, its remarkable performance is surely due to its vast generalization capabilities. In this study, performance of SAM in evaluated on 20% hold out data of Montgomery and Shenzhen dataset. SAM's performance on these completely unseen data is presented in table 1. It is found that, ViT large showed slightly better performance compared to other two versions of SAM.

Table 1. Performance of SAM with varied ViT backbone.

| Study | Dataset | Model | Dice | IoU | Kappa |
|---|---|---|---|---|---|
| Our study | Montgomery | SAM (ViT-h) | 93.5 | 91.6 | 91.2 |
| | | SAM (ViT-l) | **95.5** | **94.5** | **94** |
| | | SAM (ViT-b) | 92 | 90.2 | 89.5 |
| | Shenzhen | SAM (ViT-h) | 91.3 | 89.3 | 88 |
| | | SAM (ViT-l) | **94.9** | **93.4** | **92.6** |
| | | SAM (ViT-b) | 91.4 | 89.5 | 88 |
| [17] | Shenzhen | MWG-UNet | 85.18 | 81.36 | |
| [13] | Montgomery | Improved UNet | 97.82 | 95.55 | |
| [7] | Montgomery | ResUNet++ | 96.36 | 94.17 | |
| [19] | Montgomery | SAM (ViT-l) | 93.19 | 87.45 | |
| [21] | Montgomery | SAM (ViT-b) | 97.16 | 94.53 | |
| | Shenzhen | | 95.25 | 91.07 | |

SAM predicted lung segments with very high dice score for most of the images in the hold out dataset. But for some images (less than 5%) it performed quite poorly. Performance metrices showed in table 1 is calculated without considering these poorly performed images. In figure 9, some images for which SAM gave excellent segmentation results is shown. Where it is clearly out performing U-Net model with a high margin. In figure 10, images for which SAM gave worst results is presented. In both images 10(a) and 10(e), the left lung region is different from normal lung. There are high degree of anomaly present in both cases, which is either for severe thorax disease or for some distortion in the anatomical structure.

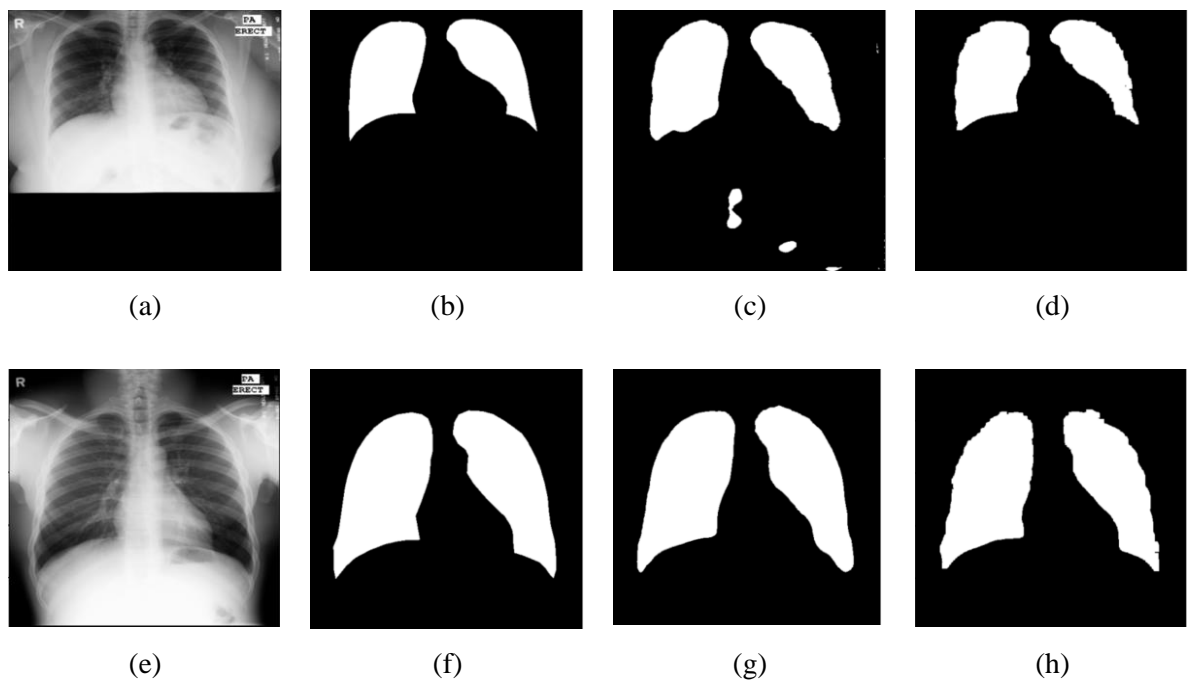

| (a) | (b) | (c) | (d) |
| (e) | (f) | (g) | (h) |

Fig. 9. SAM's segmentation best results. (a) (e) CXR images, (b) (f) Ground truth masks, (c) (g) Segmentation from UNet with pre-trained VGG19, (d) (h) Segmentation results from SAM.

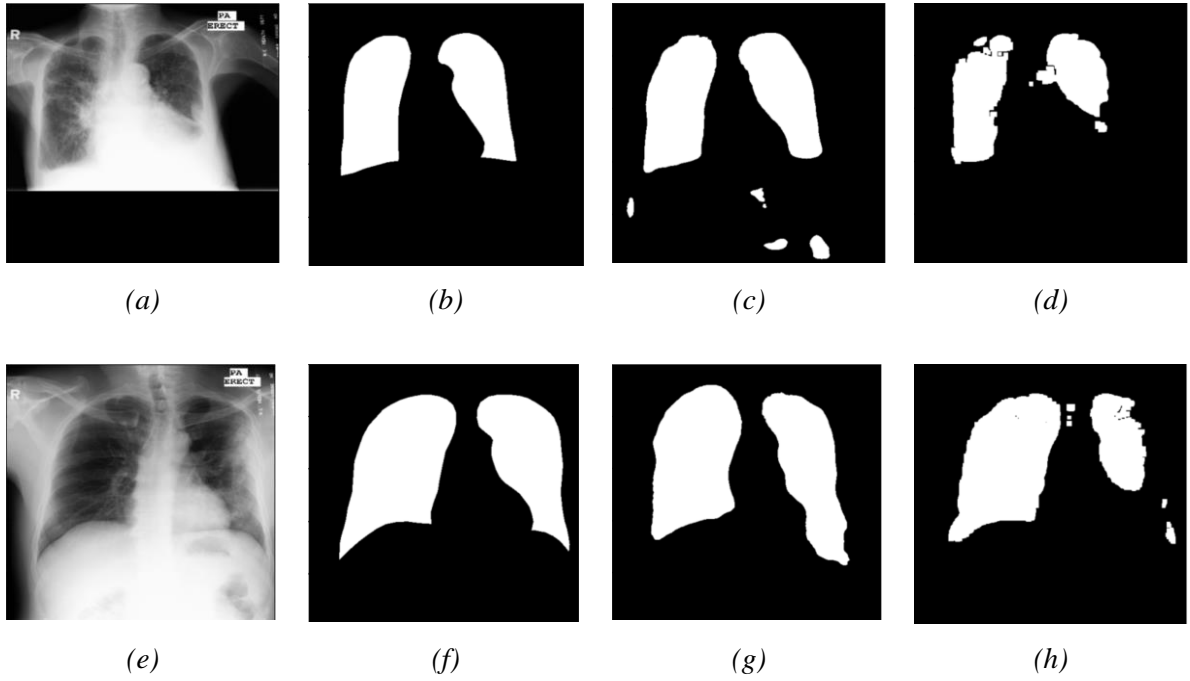

Fig. 10. SAM's segmentation worst results. (a) (e) CXR images, (b) (f) Ground truth masks, (c) (g) Segmentation from UNet with pre-trained VGG19, (d) (h) Segmentation results from SAM.

## V. Acknowledgments

This work was supported by Research Grant No. 2023-IRG-ENIT-36 under the Deanship of Research and Graduate Studies at Ajman University.